# Kalman Filtering of Distributed Time Series

**Dan Stefanoiu, Janetta Culita**

*"Politehnica" University of Bucharest, Dept. of Automatic Control and Computer Science*
*313 Splaiul Independentei 060032, Bucharest, ROMANIA*
E-mails: danny@indinf.pub.ro, jculita@yahoo.com

**Abstract:** This paper aims to introduce an application to Kalman Filtering Theory, which is rather unconventional. Recent experiments have shown that many natural phenomena, especially from ecology or meteorology, could be monitored and predicted more accurately when accounting their evolution over some geographical area. Thus, the signals they provide are gathered together into a collection of distributed time series. Despite the common sense, such time series are more or less correlated each other. Instead of processing each time series independently, their collection can constitute the set of measurable states provided by some open system. Modeling and predicting the system states can take benefit from the family of Kalman filtering algorithms. The article describes an adaptation of basic Kalman filter to the context of distributed signals collections and completes with an application coming from Meteorology.

## 1. INTRODUCTION AND PROBLEM STATEMENT

Kalman Filtering (KF) Theory was originated at early '60s by the works of R.E. Kalman and R.S. Bucy (Kalman, 1960, Kalman-Bucy, 1961). One can say that Kalman-Bucy's approach acted like a switch within the scientific community, because, nowadays, the literature on this topic is one of the richest, concerning the theory, as well as the applications. Moreover, new and sometimes surprising applications continue to keep the KF field alive. For example, one can mention the latest results from avionics (the stellar inertial navigation problem) (Kayton, 1997), fault diagnosis (Hajiyev, 2003) or robotics (Negenborn, 2003). This paper focuses on the problem of correlated time series prediction. Evolution of some natural phenomena can be monitored with higher accuracy if the observation and measurement take into account not only time variation of some parameter, but also its distribution over a geographical area. Take for example the monitoring of minimum and maximum temperatures over a geographical area (see Fig. 1).

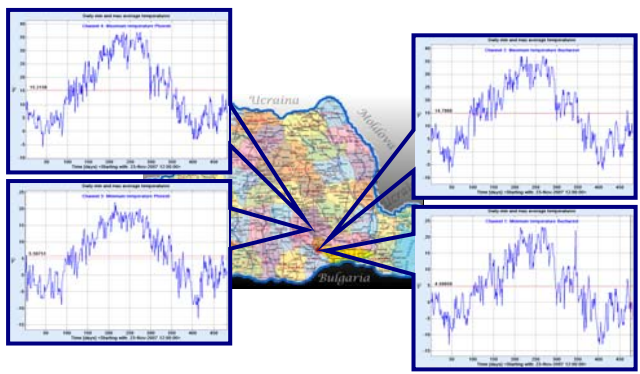

Fig. 1. Temperature monitoring in 2 cities from Romania.

When using a stand alone sensor, there is a problem with its location. Obviously, the temperature varies both in time and space. A small network of sensors is seemingly more suitable. Sensors could thus provide several time series, on different locations. Such data, coming from different channels, are in general more or less correlated. For example, in Fig. 1, one can easily notice the strong correlations between the fourth temperature variations, since the two cities are close to each-other. It is even possible that hidden correlations (that cannot be perceived) be crucial for monitoring. Assume that the monitoring goal is to predict the temperature. It is very likely that better prediction results be obtained when considering the collection of all data sets, rather than when building independent prediction models for each channel in isolation. The problem is then to build and estimate multi-variable identification models, in view of prediction.

The solution introduced within this paper relies on the idea that sensors provide direct noisy data from the states of an open and quasi-ubiquitous system. The system has in fact a continuous collection of variable states. Placing a finite set of sensors at different locations, in order to perform measuring, is equivalent to sampling the system both in time and space. The prediction problem of each data set (a time series, in fact) is actually a problem of state prediction and can thus be solved in context of KF Theory. Therefore, an adaptation of basic KF algorithm to the context of distributed time series is presented next. The article is structured as follows. Next sections introduce the Markov-Kalman-Bucy method and summarize the algorithm that allow the distributed prediction. The simulation case study is based on the example in Fig.1. A conclusion and the references list complete the paper.

## 2. DISTRIBUTED PREDICTION FRAMEWORK

The distributed prediction relies on state representation below:

$$\begin{cases} \mathbf{x}[k+1] = \mathbf{A}_k \mathbf{x}[k] + \mathbf{B}_k^x \mathbf{u}[k] + \mathbf{F}_k \mathbf{w}[k] \\ \mathbf{y}[k] = \mathbf{C}_k \mathbf{x}[k] + \mathbf{B}_k^y \mathbf{u}[k] + \mathbf{D}_k \mathbf{v}[k] \end{cases}, \quad \forall k \in \mathbb{N}, \quad (1)$$

where:

- $\mathbf{A}_k \in \mathbb{R}^{nx \times nx}$, $\mathbf{B}_k^{\mathbf{x}} \in \mathbb{R}^{nx \times nu}$, $\mathbf{B}_k^{\mathbf{y}} \in \mathbb{R}^{ny \times nu}$, $\mathbf{C}_k \in \mathbb{R}^{ny \times nx}$, $\mathbf{D}_k \in \mathbb{R}^{ny \times ny}$ and $\mathbf{F}_k \in \mathbb{R}^{nx \times nx}$ are matrices including all variable (but already estimated) parameters of some stochastic process; (usually, $\mathbf{D}_k = \mathbf{I}_{ny}$);
- $\mathbf{x} \in \mathbb{R}^{nx}$ is the unknown state vector;
- $\mathbf{u} \in \mathbb{R}^{nu}$ is the vector of (measurable) input signals;
- $\mathbf{y} \in \mathbb{R}^{ny}$ is the vector of (measurable) output signals;
- $\mathbf{w} \in \mathbb{R}^{nx}$ is the (unknown) endogenous system noise;
- $\mathbf{v} \in \mathbb{R}^{ny}$ is the (unknown) exogenous noise, which is usually corrupting the measured data.

Whenever the stochastic process cannot be stimulated artificially (like in case of time series), the input vector is assigned to null values. Therefore, the noise $\mathbf{w}$ becomes the virtual useful input, while the noise $\mathbf{v}$ is parasite.

The following noise hypotheses are assumed for model (1): (a) all noises are zero mean, Gaussian; (b) the two noises are uncorrelated each other; (c) the endogenous noise is non auto-correlated, but its compounds could be correlated at the same instant; (d) the compounds of exogenous noise are white and uncorrelated each-other. According to the last two hypotheses, the covariance matrices of noises are mostly null, excepting for the current instant $k \in \mathbb{N}$, when they are denoted by:

$$\mathbf{\Psi}_{\mathbf{w}}[k] = E\{\mathbf{w}[k]\mathbf{w}^T[k]\} \text{ and } \mathbf{\Psi}_{\mathbf{v}}[k] = E\{\mathbf{v}[k]\mathbf{v}^T[k]\}. \qquad (2)$$

Obviously, the matrix $\mathbf{\Psi}_{\mathbf{v}}$ is diagonal. This is perhaps the most restrictive condition, in general case.

In case of distributed time series, the output vector $\mathbf{y}$ includes all the data sets (on channels), whereas the state vector $\mathbf{x}$ encodes the invisible correlations between them. Naturally, there is no reason to consider that the white noises corrupting the data are correlated each other. The number of states is not necessarily equal to the number of time series. It depends on the size of multi-dimensional ARMA(X) model assigned to the global process, as shown in next section.

Two main problems can be stated within this context. The first one is to identify a rough ARMA model (by using the time series), to improve its accuracy and to transform it into a minimal state representation. The second problem is to predict the states via an adapted version of Kalman filtering. Each problem is approached next.

### 3. FROM ARMA(X) TO STATE REPRESENTATION

The rough ARMA model can be constructed by simply considering that the time series are independent each other. Thus, for each time series $y_j$ ($j \in \overline{1, ny}$) of length $N \in \mathbb{N}^*$, the corresponding ARMA model:

$$A_j(q^{-1}) y_j \equiv C_j(q^{-1}) e_j, \qquad (3)$$

(with known notations), is identified via Minimum Prediction Error Method (MPEM) (Söderström and Stoica, 1989). After identification, the estimated polynomials $\hat{A}_j$ and $\hat{C}_j$ (for $j \in \overline{1, ny}$) yield the evaluation of prediction error:

$$\hat{v}_j \equiv \hat{A}_j(q^{-1}) y_j + \left[1 - \hat{C}_j(q^{-1})\right] \hat{e}_j, \qquad (4)$$

which actually stands for the (approximated) input $u_j$ of overall stochastic process. So, $nu = ny$, in this case.

To refine the rough model, one adopts a global model, of ARMAX type:

$$\mathbf{A}(q^{-1}) \mathbf{y} \equiv \mathbf{B}(q^{-1}) \hat{\mathbf{v}} + \mathbf{C}(q^{-1}) \mathbf{e}, \qquad (5)$$

where the polynomial matrices $\mathbf{A}, \mathbf{B}, \mathbf{C} \in \mathbb{R}^{ny \times ny}(q^{-1})$ have to be identified from the time series as output data and rough prediction errors (4) as input data. The two system functions of model (5), i.e.:

$$\mathbf{H}(q^{-1}) = \mathbf{A}^{-1}(q^{-1}) \mathbf{B}(q^{-1}) \ \& \ \mathbf{G}(q^{-1}) = \mathbf{A}^{-1}(q^{-1}) \mathbf{C}(q^{-1}), \quad (6)$$

encode moreover the correlations between time series. Identification makes use of the same MPEM, but applied to a multi-dimensional stochastic process. Implementation of such a method is non trivial and involves many numerical problems. In order to reach for a suitable tradeoff between speed and accuracy, some simplifications are necessary. For example, in MATLAB environment, the following general principle has been adopted: each output depends on the input signals and noises only. With another words, output signals are not mixed each other, which implies the matrix $\mathbf{A}(q^{-1})$ is diagonal. Another simplification is related to inputs: since input signals are actually noises, the channel $j$ should not account twice the input $u_j$ (one for input and another one for noise). Consequently, the matrix $\mathbf{B}(q^{-1})$ has null diagonal. Although the resulting model is seemingly less accurate than the one obtained by applying MPEM at once, it is assumed that the accuracy is acceptable. The great facility of such an approach is that the system functions (7) are directly computed, without inverting the polynomial matrix $\mathbf{A}$.

Once the system functions being estimated, the MIMO model can be converted into a state representation like (1), following at least three rationales. First of them is based on Theorem of Division with Reminder and atomic ratios decomposition (for polynomials), following the idea introduced in (Proakis, 1996).) The second one starts from the linear regression form of ARMAX model and defines the state by concatenating the regressors vector and the parameters vector (Niedźwiecki, 2000). The number of states can increase very fast in case of MIMO models. This is the reason the third conversion technique introduced in (vanOverschee, 1996) is often preferred, since it leads to the minimum state representation.

## 4. ADAPTED KALMAN FILTER (PREDICTOR)

The filter is aiming to predict the states of model (1), by using a numerical procedure that relies on the recursive Markov estimator (Stefanoiu, 2005). Hereafter, the Markov estimator and the filter equations are described.

The main result of Gauss-Markov Theorem (GMT) (Placket, 1950) can be implemented through a recursive procedure. One starts from the canonic form of linear regression associated to an identification model (like (3) or (5)):

$$\mathbf{Y} = \mathbf{\Phi}\mathbf{\theta}^* + \mathbf{V}, \quad E\{\mathbf{V}\mathbf{V}^T\} = \mathbf{\Psi} > \mathbf{0}, \tag{7}$$

where $\mathbf{Y} \in \mathbb{R}^N$ is the $N$-length output (measurable) data vector, $\mathbf{\Phi} \in \mathbb{R}^{N \times n\theta}$ is the matrix of *regressors* (either measurable or not), $\mathbf{\theta}^* \in \mathbb{R}^{n\theta}$ is the vector of $n\theta$ unknown true parameters and $\mathbf{V} \in \mathbb{R}^N$ is the vector of (non measurable) noise data with $\mathbf{\Psi} \in \mathbb{R}^{N \times N}$ as covariance matrix. One assumes the noise is Gaussian with zero mean, but not necessarily white (i.e. the matrix $\mathbf{\Psi}$ could be non diagonal). The GMT states that the following Markov estimation:

$$\hat{\mathbf{\theta}} = \left(\mathbf{\Phi}^T \mathbf{\Psi}^{-1} \mathbf{\Phi}\right)^{-1} \mathbf{\Phi}^T \mathbf{\Psi}^{-1} \mathbf{Y} \tag{8}$$

is unbiased, consistent and efficient. Moreover, the estimation accuracy is also provided (the inverse of estimation error covariance matrix):

$$\hat{\mathbf{P}}^{-1} = \left[ E\left\{ (\hat{\mathbf{\theta}} - \mathbf{\theta}^*)(\hat{\mathbf{\theta}} - \mathbf{\theta}^*)^T \right\} \right]^{-1} = \mathbf{\Phi}^T \mathbf{\Psi}^{-1} \mathbf{\Phi}. \tag{9}$$

The problem is to design an efficient numerical procedure to compute (8) and (9) recursively. Since not only $\mathbf{\Psi}$ is usually unknown, but its size could be quite big, computing its inverse is non trivial. To solve the problem, one extends the canonical model (7) by the following virtual model (an identity, also known as *random walk model*):

$$\tilde{\mathbf{\theta}} = \mathbf{\theta}^* + \underbrace{(\tilde{\mathbf{\theta}} - \mathbf{\theta}^*)}_{\mathbf{W}}, \tag{10}$$

where $\tilde{\mathbf{\theta}} \in \mathbb{R}^{n\theta}$ is the Markov estimation from previous computation stage, whereas $\mathbf{W} \in \mathbb{R}^{n\theta}$ is the virtual noise (unknown, since $\mathbf{\theta}^*$ is non measurable). Interestingly, the covariance matrix corresponding to virtual noise, $\tilde{\mathbf{P}}$, is already available from the previous computation too. After fusing equations (7) and (10), the extended model is:

$$\underbrace{\begin{bmatrix} \mathbf{Y} \\ \tilde{\mathbf{\theta}} \end{bmatrix}}_{\tilde{\mathbf{Y}}} = \underbrace{\begin{bmatrix} \mathbf{\Phi} \\ \mathbf{I} \end{bmatrix}}_{\tilde{\mathbf{\Phi}}} \mathbf{\theta}^* + \underbrace{\begin{bmatrix} \mathbf{V} \\ \mathbf{W} \end{bmatrix}}_{\tilde{\mathbf{V}}}, \tag{11}$$

with natural notations. Whenever the noise $\mathbf{V}$ is not correlated to the virtual noise $\mathbf{W}$, the covariance matrix of extended noise $\tilde{\mathbf{V}}$, includes two diagonal blocks only:

$$\tilde{\mathbf{\Psi}} = E\{\tilde{\mathbf{V}} \cdot \tilde{\mathbf{V}}^T\} = \begin{bmatrix} \mathbf{\Psi} & \mathbf{0} \\ \mathbf{0} & \tilde{\mathbf{P}} \end{bmatrix}. \tag{12}$$

When computing again the Markov estimation with the extended stochastic process (11), after some manipulations, the direct link between the current and previous values of estimated parameters can be revealed:

$$\hat{\mathbf{\theta}} = \tilde{\mathbf{\theta}} + \underbrace{\left(\mathbf{\Phi}^T \mathbf{\Psi}^{-1} \mathbf{\Phi} + \tilde{\mathbf{P}}^{-1}\right)^{-1} \mathbf{\Phi}^T \mathbf{\Psi}^{-1}}_{\mathbf{\Gamma}} \underbrace{(\mathbf{Y} - \mathbf{\Phi}\tilde{\mathbf{\theta}})}_{\tilde{\mathbf{\varepsilon}}}. \tag{13}$$

Two interesting terms are outlined in (13). The first one is the *prediction error* $\tilde{\mathbf{\varepsilon}} \in \mathbb{R}^N$, based on previous estimation $\tilde{\mathbf{\theta}}$. The second one is the *sensitivity gain* $\mathbf{\Gamma} \in \mathbb{R}^{n\theta \times N}$, based on previous accuracy $\tilde{\mathbf{P}}^{-1}$. Supplementary manipulations can lead to an equivalent expression of the gain, which is more suitable for numerical evaluations (only one matrix has to be inverted instead of 4):

$$\mathbf{\Gamma} = \tilde{\mathbf{P}} \mathbf{\Phi}^T \left(\mathbf{\Psi} + \mathbf{\Phi} \tilde{\mathbf{P}} \mathbf{\Phi}^T\right)^{-1}. \tag{14}$$

The covariance matrix of estimation error can also be updated without any matrix inversion (after computing the gain):

$$\hat{\mathbf{P}} = \left(\tilde{\mathbf{P}}^{-1} + \mathbf{\Phi}^T \mathbf{\Psi}^{-1} \mathbf{\Phi}\right)^{-1} = \left(\mathbf{I}_{n\theta} - \mathbf{\Gamma}\mathbf{\Phi}\right) \tilde{\mathbf{P}}, \tag{15}$$

where $\mathbf{I}_{n\theta}$ is the unit matrix of size $n\theta$. Practically, the recursive equations (14), (15) and (13) constitute de core of implementation procedure for Markov estimator. An estimation of $\mathbf{\Psi}$ can be obtained by using the prediction error $\tilde{\mathbf{\varepsilon}}$ instead of noise $\mathbf{V}$ (since, actually, $\tilde{\mathbf{\varepsilon}}$ is the current estimation of $\mathbf{V}$.) The most time consuming operation is computing the gain (14), because, at every step, a $N \times N$ matrix has to be inverted. To reduce the computational effort, the number of adaptation data should be used instead of whole data number. So, $N$ is the number of data between successive parameters upgrading (usually, no bigger than 10).

Recall the state representation (1) and assume that all parameters are already known at current instant $k \in \mathbb{N}$. Then, the final goal is to estimate/predict the state values at the next instant $k+1$, i.e. $\hat{\mathbf{x}}[k+1]$, depending on current state values

$\hat{\mathbf{x}}[k]$ and newly measured data. In order to reach for this goal, Markov estimator has to be employed as main tool.

The main variables of Markov estimator are identified as follows from the last equation of model (1): the state vector $\mathbf{x}[k]$ is $\boldsymbol{\theta}^*$ and the output matrix $\mathbf{C}_k$ is $\boldsymbol{\Phi}$. Consequently, $\hat{\mathbf{x}}[k]$ is $\tilde{\boldsymbol{\theta}}$, while $\hat{\mathbf{x}}[k+1]$ stands for $\hat{\boldsymbol{\theta}}$. The covariance matrix from (9) is defined as follows:

$$\hat{\mathbf{P}}_k \stackrel{\text{def}}{=} E\{(\hat{\mathbf{x}}[k]-\mathbf{x}[k])(\hat{\mathbf{x}}[k]-\mathbf{x}[k])^T\}, \quad \forall k \in \mathbb{N}. \quad (16)$$

Before starting the Markov estimation procedure, it is necessary to verify whether the noise $\mathbf{v}[k]$ is correlated to the error $(\hat{\mathbf{x}}[k]-\mathbf{x}[k])$ or not. Since the current state $\mathbf{x}[k]$ only depends on inside noise $\mathbf{w}$ and the estimated state $\hat{\mathbf{x}}[k]$ is determined by the former state $\hat{\mathbf{x}}[k-1]$ (i.e. by the values of noises at instant $k-1$ at most), the non correlation restriction is fully verified (under the noise hypotheses of section 2). This allows the current state to roughly be estimated as follows:

$$\begin{cases} \boldsymbol{\Gamma}_k = \hat{\mathbf{P}}_k \mathbf{C}_k^T \left(\mathbf{D}_k \boldsymbol{\Psi}_\mathbf{v}[k]\mathbf{D}_k^T + \mathbf{C}_k \hat{\mathbf{P}}_k \mathbf{C}_k^T\right)^{-1} \\ \tilde{\mathbf{P}}_k = \hat{\mathbf{P}}_k - \boldsymbol{\Gamma}_k \mathbf{C}_k \hat{\mathbf{P}}_k \\ \tilde{\mathbf{x}}[k] = \hat{\mathbf{x}}[k] + \boldsymbol{\Gamma}_k \left(\mathbf{y}[k] - \mathbf{C}_k \hat{\mathbf{x}}[k] - \mathbf{B}_k^\mathbf{y} \mathbf{u}[k]\right) \end{cases}, \quad \forall k \in \mathbb{N}, \quad (17)$$

since the exogenous noises are mixed through matrix $\mathbf{D}_k$. In (17), $\tilde{\mathbf{x}}[k]$ is not yet equal to $\hat{\mathbf{x}}[k+1]$ and $\tilde{\mathbf{P}}_k$ is different from $\hat{\mathbf{P}}_{k+1}$ as well, because the first equation of model (1) remained untouched. They are only rough approximations of the targeted terms. To refine the approximations, the next state is computed from the first equation of model (1), with $\tilde{\mathbf{x}}[k]$ instead of $\hat{\mathbf{x}}[k]$ and $\mathbf{w}=\mathbf{0}$. Thus:

$$\hat{\mathbf{x}}[k+1] = \mathbf{A}_k \tilde{\mathbf{x}}[k] + \mathbf{B}_k \mathbf{u}[k], \quad \forall k \in \mathbb{N}. \quad (18)$$

The next covariance matrix $\hat{\mathbf{P}}_{k+1}$ results then as direct consequence of equation (18):

$$\hat{\mathbf{P}}_{k+1} \stackrel{\text{def}}{=} \mathbf{A}_k \tilde{\mathbf{P}}_k \mathbf{A}_k^T + \mathbf{F}_k \boldsymbol{\Psi}_\mathbf{w}[k]\mathbf{F}_k^T, \quad \forall k \in \mathbb{N}, \quad (19)$$

after some manipulations where the non correlation between $\mathbf{w}[k]$ and $(\tilde{\mathbf{x}}[k]-\mathbf{x}[k])$ played the main role. After aggregating equations (17)-(19), the kernel of final recursive procedure related to Kalman-Bucy filtering is obtained:

$$\begin{cases} \boldsymbol{\Gamma}_k = \hat{\mathbf{P}}_k \mathbf{C}_k^T \left(\mathbf{D}_k \boldsymbol{\Psi}_\mathbf{v}[k]\mathbf{D}_k^T + \mathbf{C}_k \hat{\mathbf{P}}_k \mathbf{C}_k^T\right)^{-1} \\ \hat{\mathbf{x}}[k+1] = \mathbf{A}_k \hat{\mathbf{x}}[k] + \mathbf{B}_k^\mathbf{x} \mathbf{u}[k] + \mathbf{A}_k \boldsymbol{\Gamma}_k \left(\mathbf{y}[k] - \mathbf{C}_k \hat{\mathbf{x}}[k] - \mathbf{B}_k^\mathbf{y} \mathbf{u}[k]\right), \\ \hat{\mathbf{P}}_{k+1} = \mathbf{F}_k \boldsymbol{\Psi}_\mathbf{w}[k]\mathbf{F}_k^T + \mathbf{A}_k \left(\hat{\mathbf{P}}_k - \boldsymbol{\Gamma}_k \mathbf{C}_k \hat{\mathbf{P}}_k\right)\mathbf{A}_k^T \end{cases}$$
$$\forall k \in \mathbb{N}. \quad (20)$$

Obviously, the procedure (20) can be implemented only after estimating the mixed covariance matrices of noises, $\mathbf{D}_k \boldsymbol{\Psi}_\mathbf{v}[k]\mathbf{D}_k^T$ and $\mathbf{F}_k \boldsymbol{\Psi}_\mathbf{w}[k]\mathbf{F}_k^T$ at any instant $k \in \mathbb{N}$. This operation involves 2 computation stages. At the first step, the mixed exogenous noise is estimated with the help of the second equation of (1).

$$\mathbf{D}_n \hat{\mathbf{v}}[n] = \boldsymbol{\varepsilon}[n] = \mathbf{y}[n] - \mathbf{C}_n \hat{\mathbf{x}}[n] - \mathbf{B}_n^\mathbf{y} \mathbf{u}[n], \quad \forall n \in \overline{0,k}. \quad (21)$$

At the second step, the covariance matrix is updated:

$$\mathbf{D}_k \boldsymbol{\Psi}_{\hat{\mathbf{v}}}[k]\mathbf{D}_k^T = \frac{1}{k+1}\left(\sum_{n=0}^{k}\mathbf{D}_n \hat{\mathbf{v}}[n]\hat{\mathbf{v}}^T[n]\mathbf{D}_n^T\right) = \\ = \frac{1}{k+1}\left(k\mathbf{D}_{k-1}\boldsymbol{\Psi}_{\hat{\mathbf{v}}}[k-1]\mathbf{D}_{k-1}^T + \mathbf{D}_k \hat{\mathbf{v}}[k]\hat{\mathbf{v}}^T[k]\mathbf{D}_k^T\right). \quad (22)$$

For the endogenous noise, instead of repeating the steps above, one can compute the estimation more elegantly. Thus, it is easy to see from (20), (21) and (1) that:

$$\mathbf{A}_k \boldsymbol{\Gamma}_k \mathbf{D}_k \hat{\mathbf{v}}[k] = \hat{\mathbf{x}}[k+1] - \mathbf{A}_k \hat{\mathbf{x}}[k] - \mathbf{B}_k^\mathbf{x} \mathbf{u}[k] = \mathbf{F}_k \hat{\mathbf{w}}[k]. \quad (23)$$

Since the noises estimations are so correlated, the estimation of $\mathbf{F}_k \boldsymbol{\Psi}_\mathbf{w}[k]\mathbf{F}_k^T$ follows straightforwardly:

$$\mathbf{F}_k \boldsymbol{\Psi}_{\hat{\mathbf{w}}}[k]\mathbf{F}_k^T = \frac{1}{k+1}\sum_{n=0}^{k}\mathbf{F}_n \hat{\mathbf{w}}[n]\hat{\mathbf{w}}^T[n]\mathbf{F}_n^T = \\ = \frac{1}{k+1}\left(k\mathbf{F}_{k-1}\boldsymbol{\Psi}_{\hat{\mathbf{w}}}[k-1]\mathbf{F}_{k-1}^T + \mathbf{A}_k \boldsymbol{\Gamma}_k \mathbf{D}_k \hat{\mathbf{v}}[k]\hat{\mathbf{v}}^T[k]\mathbf{D}_k^T \boldsymbol{\Gamma}_k^T \mathbf{A}_k^T\right). \quad (25)$$

Equations above can be gathered together into a numerical recipe aiming to predict the discrete stochastic states and outputs of model (1). The main steps are as follows.

➤ *Input data: a small collection of time series values (the training set $\mathcal{D}_0 = \{\mathbf{y}[n]\}_{n \in \overline{1,N_0}}$) yielding initialization.*

1. *Initialization. Produce the first state representation (1). Then complete the initialization by setting: an arbitrary state vector $\hat{\mathbf{x}}_0$, the covariance matrices $\hat{\mathbf{P}}_0 = \alpha \mathbf{I}_{nx}$ (with $\alpha \in \mathbb{R}_+^*$), $\mathbf{F}_{-1}\boldsymbol{\Psi}_{\hat{\mathbf{w}}}[-1]\mathbf{F}_{-1}^T = \mathbf{0}_{nx}$ and $\mathbf{D}_{-1}\boldsymbol{\Psi}_{\hat{\mathbf{v}}}[-1]\mathbf{D}_{-1}^T = \mathbf{0}_{ny}$.*

2. *For $k \geq 0$:*
   2.1. *Estimate the exogenous mixed noise:*
   $$\mathbf{D}_k \hat{\mathbf{v}}[k] = \mathbf{y}[k] - \mathbf{C}_k \hat{\mathbf{x}}[k] - \mathbf{B}_k^\mathbf{y} \mathbf{u}[k].$$
   2.2. *Update the covariance matrix of exogenous noise:*
   $$\mathbf{D}_k \boldsymbol{\Psi}_{\hat{\mathbf{v}}}[k]\mathbf{D}_k^T = \frac{1}{k+1}\left(k\mathbf{D}_{k-1}\boldsymbol{\Psi}_{\hat{\mathbf{v}}}[k-1]\mathbf{D}_{k-1}^T + \mathbf{D}_k \hat{\mathbf{v}}[k]\hat{\mathbf{v}}^T[k]\mathbf{D}_k^T\right).$$
   2.3. *Compute the auxiliary matrix: $\mathbf{Q}_k = \mathbf{C}_k \hat{\mathbf{P}}_k$.*
   2.4. *Invert the matrix: $\mathbf{R}_k = \mathbf{D}_k \boldsymbol{\Psi}_{\hat{\mathbf{v}}}[k]\mathbf{D}_k^T + \mathbf{Q}_k \mathbf{C}_k^T \in \mathbb{R}^{ny \times ny}$.*
   2.5. *Evaluate the sensitivity gain: $\boldsymbol{\Gamma}_k = \mathbf{Q}_k^T \mathbf{R}_k^{-1}$.*
   2.6. *Compute the auxiliary matrix $\mathbf{S}_k = \mathbf{A}_k \boldsymbol{\Gamma}_k$.*

*2.7. Update the covariance matrix of endogenous noise:*
$$\mathbf{F}_k \mathbf{\Psi}_{\hat{\mathbf{w}}}[k] \mathbf{F}_k^T = \frac{1}{k+1}\left(k\mathbf{F}_{k-1}\mathbf{\Psi}_{\hat{\mathbf{w}}}[k-1]\mathbf{F}_{k-1}^T + \mathbf{S}_k \mathbf{D}_k \hat{\mathbf{v}}[k]\hat{\mathbf{v}}^T[k]\mathbf{D}_k^T \mathbf{S}_k^T\right).$$

*2.8. Update the covariance matrix of estimation error:*
$$\hat{\mathbf{P}}_{k+1} = \mathbf{F}_k \mathbf{\Psi}_{\hat{\mathbf{w}}}[k] \mathbf{F}_k^T + \mathbf{A}_k\left(\hat{\mathbf{P}}_k - \mathbf{\Gamma}_k \mathbf{Q}_k\right)\mathbf{A}_k^T.$$

*2.9. Predict the state:*
$$\hat{\mathbf{x}}[k+1] = \mathbf{A}_k \hat{\mathbf{x}}[k] + \mathbf{B}_k^{\mathbf{x}}\mathbf{u}[k] + \mathbf{S}_k \mathbf{D}_k \hat{\mathbf{v}}[k].$$

*2.10. Predict the output:* $\hat{\mathbf{y}}[k+1] = \mathbf{C}_k \hat{\mathbf{x}}[k+1] + \mathbf{B}_k^{\mathbf{y}}\mathbf{u}[k+1]$.

*2.11. Acquire new data:* $\mathcal{D}_{k+1} = \mathcal{D}_k \cup \{\mathbf{y}[k+1]\}$.

*2.12. Update the state model.*

➢ *Output data:*
- predicted time series values $\{\hat{\mathbf{y}}[k]\}_{k \in \mathbb{N}^*}$;
- estimated covariance matrices $\{\mathbf{D}_k \mathbf{\Psi}_{\hat{\mathbf{v}}}[k]\mathbf{D}_k^T\}_{k \in \mathbb{N}^*}$.

The most time consuming steps of algorithm above is 2.12 (state model matrices updating), followed by 2.4 (matrix inversion). The algorithm above can easily be adapted to multi-step prediction, thanks to Markov estimator. In this case, the algorithm has two stages. The first one is concerned with the model adaptation. In the second one, multi-step prediction is performed. The main difference between the two algorithms (one step and multi-step prediction) consists of the exogenous noise estimation. As long as the measured data are available, equation (21) can successfully be employed. When the measured data are missing, the exogenous noises have to be estimated by a different technique. For example, MIMO-ARMA(X) models can be employed in this aim; the estimated white noises can directly stand for mixed exogenous noises (since, usually, $\mathbf{D}_n = \mathbf{I}_{ny}$).

Estimations of covariance matrices $\mathbf{D}_{N_y+k}\mathbf{\Psi}_{\hat{\mathbf{v}}}[N_y+k]\mathbf{D}_{N_y+k}^T$ are necessary both to assess the prediction performance and to estimate the SNRs. Thus, the diagonal of each matrix returns the set $\{\hat{\sigma}_{j,k}^2\}_{j\in\overline{1,ny}}$, whose values play the role of prediction errors variances on every channel. Here, $N_y$ is the data length and $k \in \overline{1,K}$ is the current prediction step on the prediction horizon. Then the following two types of SNR can be evaluated (one for measured data and another one for predicted data):

$$\text{SNR}_j \stackrel{def}{=} \sigma_{y_j}^2 / \hat{\sigma}_{j,1}^2 \;\&\; \text{SNR}_j^K \stackrel{def}{=} \left(\sigma_{y_j}^K\right)^2 / \left(\sigma_{y_j - \hat{y}_j}^K\right)^2, \forall j \in \overline{1,ny} \quad (26)$$

where $\sigma_{y_j}$, $\sigma_{y_j}^K$ are standard deviations of data on measuring and prediction horizons, respectively; also, $\sigma_{y_j - \hat{y}_j}^K$ is the standard deviation of prediction error. The SNRs (26) allow one to define the *prediction quality* (PQ) cost function below:

$$\text{PQ}_j \stackrel{def}{=} 100 / \left(1 + \frac{\sqrt{\sum_{k=1}^{K}\hat{\sigma}_{j,k}^2}}{\hat{\lambda}_{e_j}\sqrt{\text{SNR}_j}\sqrt{\text{SNR}_j^K}}\right) [\%], \; \forall j \in \overline{1,ny}. \quad (27)$$

The bigger the norm of $\mathbf{PQ} = [\text{PQ}_1 \cdots \text{PQ}_{ny}]^T$, the better the predictor performance.

Finding the structural indices that maximize the norm of $\mathbf{PQ}$ cannot be realized through an exhaustive search. The structural indices are: $p$ – the degree of polynomial trend, $na$, $nc$ – the orders of ARMA model and $nx$ – the number of states for the linear system (1). An evolutionary searching technique has to be employed in this aim.

## 5. SIMULATION RESULTS

An application coming from Meteorology has been considered. Daily minimum and maximum temperatures of two cities have been monitored and predicted (as Fig.1 suggests). The cities are 60 km far each other on a plain. The data block consists of 482 samples on 4 channels. Two predictors are compared in terms of PQ: PARMA and KARMA. The first one is based on ARMA prediction of each channel in isolation. The second one relies on the adapted Kalman filter predictor. Both algorithms have been implemented within MATLAB environment. In order to find optimal structural indices, the technique from (Kennedy, 1997) has been adopted. There are many implementation details that cannot be described here. Just one word regarding KARMA: numerical stability of the algorithm required special attention at step 2.5.

For each one of the final figures (2-9), three variations are depicted: the original data with the deterministic model (trend and seasonal variation, if any) on top, the residual noise with estimated SNR in the middle and the performance on the prediction horizon at bottom. Although the predictability varies from a channel to another, KARMA succeeded to perform better than PARMA (higher PQ and SNR values). (This result was confirmed by other data blocks as well, where correlation between channels exits.) However, in general, PARMA has superior performance on data blocks with (almost) uncorrelated channels. Below, the PQ values and norms are shown:

$$\mathbf{PQ}_{ARMA}^T = [46.00 \; 62.65 \; 65.12 \; 63.23] \Rightarrow \|\mathbf{PQ}_{ARMA}\| \cong 119.50;$$
$$\mathbf{PQ}_{KARMA}^T = [49.59 \; 70.18 \; 79.47 \; 71.70] \Rightarrow \|\mathbf{PQ}_{KARMA}\| \cong 137.26.$$
(28)

Only 4 states were necessary to represent the linear system associated to data. The fact $nx = ny$ in this case is pure coincidence. On the figures corresponding to KARMA performance (Figs. 6-9), the only purpose of ARMA models is to estimate the input colored noises that stimulate the system. The data on the first channel are seemingly the less predictable. This is proven by the modest PQ values returned even by KARMA (only slightly superior to PARMA one). It seems that data from this channel are less correlated to data from the other channels, which cannot be noticed by simply inspecting the data.

The prediction accuracy has increased at the expense of computational complexity for KARMA. Therefore, if the data are quite uncorrelated across channels, PARMA should be employed as the first option.

## 6. CONCLUSION

One can say that KF is a new and old topic at the same time. Concerning the theory, KF has drawn the bottom line long time ago. The applications rejuvenate however this approach. The KF-based algorithm introduced in this article is genuine. Its major contribution consists of noises estimation during the prediction. The most KF algorithms try to avoid this problem. The simulation case study on natural data has proven that the prediction quality can be improved when considering correlations between channels.

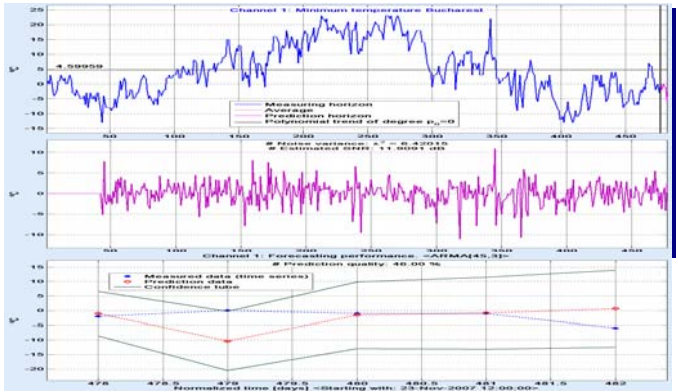

Fig. 2. PARMA performance on channel 1.

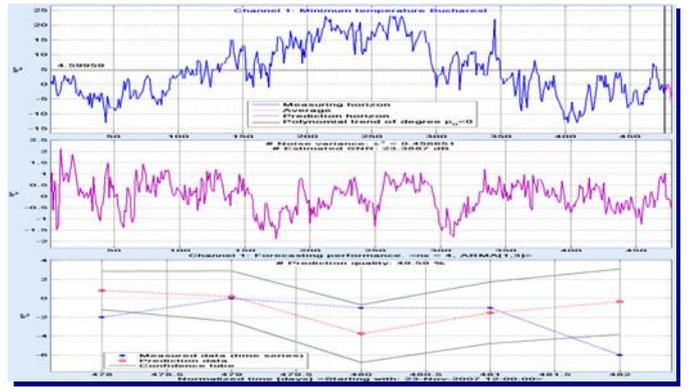

Fig. 6. KARMA performance on channel 1.

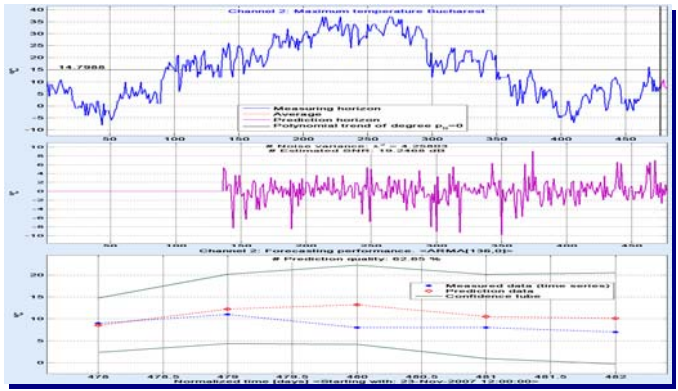

Fig. 3. PARMA performance on channel 2.

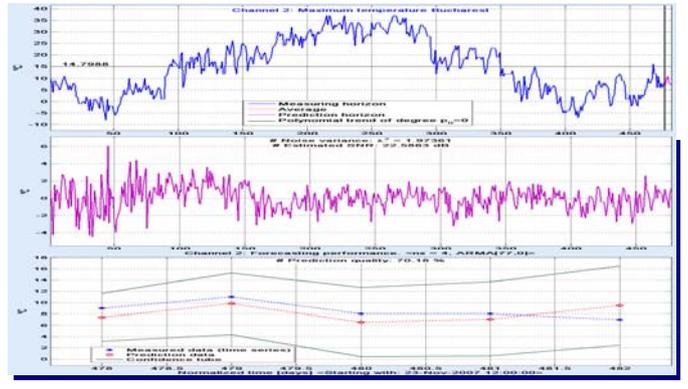

Fig. 7. KARMA performance on channel 2.

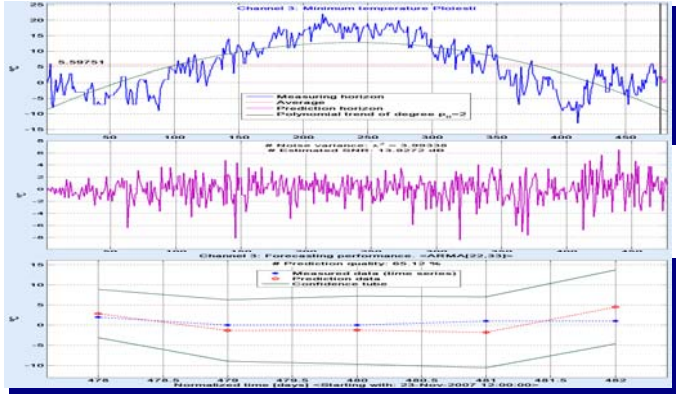

Fig. 4. PARMA performance on channel 3.

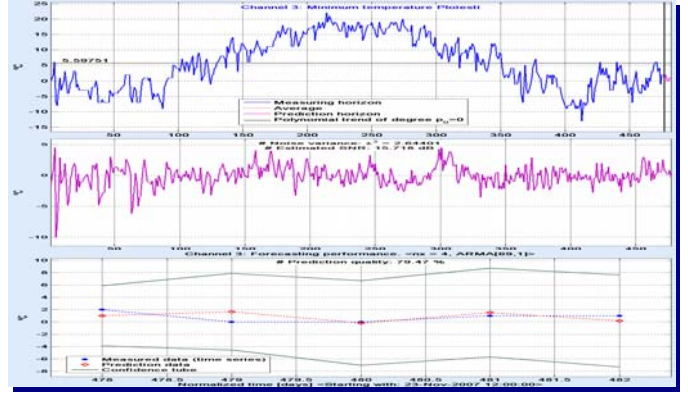

Fig. 8. KARMA performance on channel 3.

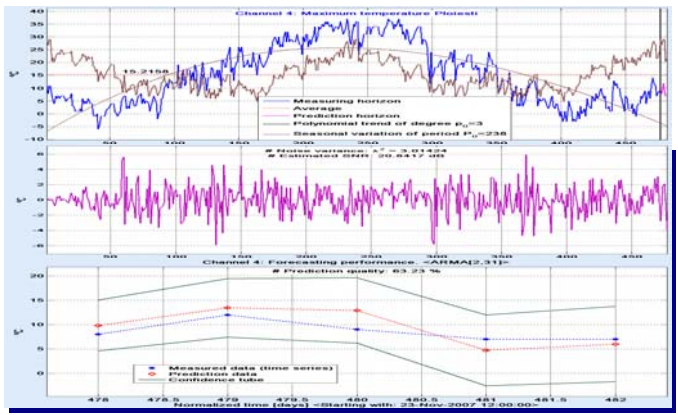

Fig. 5. PARMA performance on channel 4.

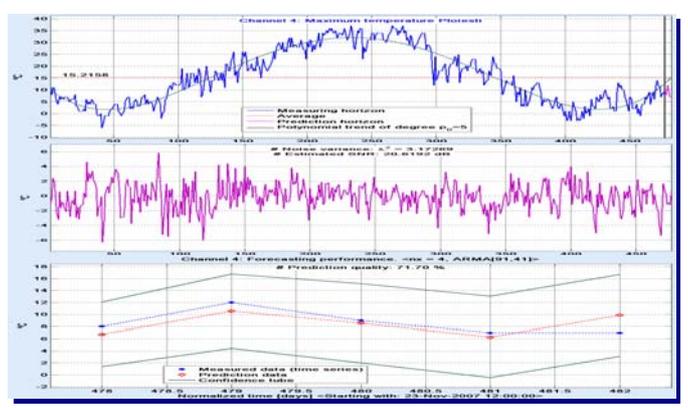

Fig. 9. KARMA performance on channel 4.